# How AI Generates Creativity from Inauthenticity

James Brusseau and
Luca Turchet

**Abstract**
Artificial creativity is presented as a counter to Benjamin's conception of an "aura" in art. Where Benjamin sees authenticity as art's critical element, generative artificial intelligence operates as pure inauthenticity. Two elements of purely inauthentic art are described: elusiveness and reflection. Elusiveness is the inability to find an origin-story for the created artwork, and reflection is the ability for perceivers to impose any origin that serves their own purposes. The paper subsequently argues that these elements widen the scope of artistic and creative potential. To illustrate, an example is developed around musical improvisation with an artificial intelligence partner. Finally, a question is raised about whether the inauthentic creativity of AI in art can be extended to human experience and our sense of our identities.

## 1 Introduction

A new kind of creative art emerges with generative technologies. This creativity flourishes in an environment of inauthenticity. It comes from no concrete place, it can be located at no particular time, and these absences are not detrimental. Instead, they are liberating. Stronger, they make an alternative kind of creativity possible. Inauthenticity is now an engine of art.

This paper shows how to reimagine art as the positive product of inauthenticity, and describes why the vision is emerging now. Artificial creativity is presented as a counter to Benjamin's conception of an "aura" in art [2]. Where Benjamin sees authenticity as art's critical element, generative artificial intelligence operates as pure inauthenticity. Two elements of purely inauthentic art are described: elusiveness and

James Brusseau
Department of Philosophy, Pace University, NYC, e-mail: jbrusseau@pace.edu

Luca Turchet
Department of Information Engineering and Computer Science, University of Trento, Italy e-mail: luca.turchet@unitn.it

reflection. Elusiveness is the inability to find an origin-story for the created artwork, and reflection is the ability for perceivers to impose any origin that serves their own purposes. In the end, the paper asks whether this new creativity is also a new way of being human.

## 2  Benjamin's structures for analyzing creative art and technology

Walter Benjamin's *Work of Art in the Age of Mechanical Reproduction* guided early twentieth century thinking about creativity intersecting with technology. The core idea is widely known and well-suited to a time when the mechanisms of massified reproduction were first spreading widely. Photography was Benjamin's paradigmatic case, but the dynamic stretched to include film, some forms of printing, the early phonograph, and the radio. What all of it did was jeopardize the origin story of artworks. These tools of reproduction broke the link between the creators' initial life, and the observers' later experience [2].

For example, the photographic reproduction of the Mona Lisa portrait brightens some tones while dampening others. These kinds of adjustments reform the elusive smile and the sternness of her brows, and, they are inseparable from the photographic mechanism. They are even intrinsic to the chemical process of developing a picture. Mechanical reproduction, consequently, splits the portrait as the painter imagined it from the way it is seen when reprinted on cards and in books.

More profoundly, the experience of the original painter disassociates from the observers' reception. The Mona Lisa canvas displays its physical age with cracking and ridges but these are only slivers of discoloration in the typical photographic reproduction, which means that a palpable sense of years is disappearing with the conversion to a printed image. Then, the loss extends to become a partial disappearance of the artist as an individual who lived at a particular time and place. What is gone are the material traces locating the artist's reality somewhere deep in the past.

What matters overall is that the artist's intentions and human experience can no longer be discerned by viewers of the reproduction. This break between the historical and cultural world of the creator, and the mass reproductions distributed to viewers is, for Benjamin, a loss of authenticity, and the deflation of what he named the "aura" of the art.

Because the problem is authenticity, it cannot be remedied technically. It will not help to make the machines more accurate. Just like an improved forgery of painting does not make the copy less of a forgery (it actually makes it more of a forgery), so too the refinement of reproductive tools does not re-inflate the lost aura. It only provides a still starker reminder that the artist did not make the work, a machine did. So, Benjamin sought authenticity. The status of the origin story is where he began analyzing when investigating the intersection of creative art with technology.

## 3 Why is there no authenticity in generative AI art?

The skills and tools of today's engineering are very different from those animating photography, film, and the phonograph [1]. Particularly – but not only – in the area of generative AI image production, the disruptions surrounding origin-stories have transformed: while it is true that technology is once again challenging authenticity, the jeopardy this time runs deeper. For Benjamin, technology separated art from its creator. In genAI, there is no historical origin to separate from: nothing has been lost or threatened by the machines because there was nothing there at the beginning.

The reason there is no origin is that there is too much origin. And, this excess is a characteristic of how generative AI works technically. When Midjourney or Stability is prompted to create an image, the AI recompiles stacks of paintings stored in its memory, and filters their pixels for features responsive to the prompt. It may be that in that process the Mona Lisa and images of the Mona Lisa, and images of images of Mona Lisa are active somewhere. But where? And to what degree? There are so many squares of color referencing each other that the idea that one or any group could stand out as significant makes no sense. Finding the authentic beginning of a generative AI image would be like trying to find the one grain of sand that first rested on a long beach. It is not an impossible task; it is an irrational one.

Of course, it is true that the data training AI generative models does come from the past, and directly or indirectly from specific material objects created by human hands. But even if it is conceded that the Mona Lisa did, in fact, play a significant role in the production of a specific genAI image, then why was it chosen? Why that painting instead of Rembrandt's Nightwatch, or one from Caravaggio's Mathew series? There are no answers to these questions. The magnitude is too high, the number of images called upon to respond to any prompt, the number of predictive decisions made at every pixel, and how that color decision affects the next pixel's tone, all of these calculations are superhuman in the sense that their size carries them beyond the capacities of human comprehension.

The result is art that comes from no origin precisely because it comes from so many. It is art without authenticity.

## 4 The loss of authenticity today, and the digital shine in genAI images

The absence of authenticity is a kind of digital shine recognizable in AI depictions and betraying their artificiality. These betrayals are not crude factual errors, they are not men with six fingers or women whose ears appear on top of their long hair. These are superficial mistakes, and possibly correctable with still more training data and algorithmic refinement. The digital shine is a different concept. It results not from objective mistakes so much as two aspects of AI creativity: elusiveness, and reflection.

Elusiveness is the impossibility of defining the image's historical context – there is no way to discern exactly where the image came from culturally. When Midjourney or Stable Diffusion are prompted, the resulting image belongs anywhere and nowhere in art history.

This particular elusiveness in images is not new. It can be seen on European banknotes. For anyone old enough to remember, one of the more palpable excitements of the introduction of the monetary aspect of the European Union was the introduction of new notes. On the morning of their release, people lined up at ATM machines to get their bills. The disappointment was universal. The images on the banknotes seemed insipid and empty. All of them had been carefully designed to feel very European, but even more importantly, to not evoke any particular European place. So, there were Romanesque bridges pictured on the banknotes, but not a particular Romanesque bridge in Spain or Paris. And, there were colored glass windows from cathedrals, but not from any particular cathedral, not one in Germany or Prague. Everything on the bills felt generic and was generic. Which was intentional. The message was that every bill belonged equally to every city and nation in Europe. The effacement of particularity was a political strategy.

The strategy works. Looking at the bills, there are depictions of common buildings and constructions, but it is impossible to tell exactly which building or which construction is being depicted. The original physical objects are infinitely elusive.

The elusiveness makes sense for money and politics, but up to now, it has not been reasonable - or even considered possible – for art. GenAI images make it possible. The data and algorithmic outputs present the same anonymity as the European banknotes, the anonymity which is not just the absence of an origin-story but, more significantly, the absence of a desire for an origin story. GenAI art is not faking anything or concealing anything or covering anything. Like the images on European banknotes, the contents of the productions unmistakably announce their cultural ambiguity, and no matter how hard you try to track down any authenticity – any single, human origin-story for an image – it will escape you. This is not a detail about the art (or the banknotes), it is the essence.

Besides projecting elusiveness, the digital shine of AI art is also reflective. The reflection is of the viewers themselves. Where the art of Benjamin's aura is understood as a window looking back into the past and at the creators' performing their labors, genAI art reflects the observers' projects and intentions back at them. Observers see only their own experiences and future in the art that they summon. When prompting a generative AI platform, the goal is to produce an image that is useful, that relates to what the prompter wants to achieve. The entire meaning of the digital output is contained within the aspirations of those who are doing the prompting. Analogously, in the European banknotes picturing an aqueduct or a cathedral facade, any viewer can reasonably decide that the stone construction on the bill replicates the particular stone construction they happen to know from their own city, or the one they viewed on a trip they have taken. The question is not about the image's true source, but how the viewer can use the image now.

Regardless of whether the subject is genAI or European banknotes, what matters is the form of the fundamental question elicited by the image. It is not: "Where did it come from?" Instead, it is "What can I do with it?" and "How does it serve my projects?"

So, there is a kind of inauthenticity emerging from AI today, and this is not inauthenticity as we understood that word previously. It is not inauthenticity as lack or as incomplete authenticity. Instead, this is a kind of counter-authenticity. It occupies its own positive reality, and offers a distinct structure for creative production. Where traditional authenticity precedes the beginning of artistic creation, inauthenticity comes after the generation. And, where authenticity expresses the original artist, inauthenticity is a projection of the observer. And, while authenticity transmits through the work to its eventual viewers, inauthenticity is imposed onto the work and reflects whatever origin story the viewer decides to imbue in the image.

More broadly, instead of authenticity that reveals a truth about the past, inauthenticity points a way into the future. Instead of authenticity belonging to a lost creator, inauthenticity guides an existing perceiver. Instead of authenticity being something in the art, inauthenticity is now something that is done to the art.

Summarizing, the digital shine is the genAI counter to Benjamin's aura. It is composed of elusiveness and reflection. The elusiveness is the inability to find any origin story in the work. The reflection is the invitation to viewers to see their own conception of an origin – and future – in the work.

## 5 Example of how the purpose of a work comes after the creation, from music: Nietzsche's a-tonal piano compositions

In tonal music, the key is the set of notes coming before the composition and guiding what notes may, and may not be played [3]. Atonal compositions are different, nothing is out of bounds, not even inharmonious and clashing sounds [4]. One result is that atonal music requires the listener to impose order on the sounds after they are created. It is the listener more than the composer who creates order from noise. One example is Friedrich Nietzsche's Manfred-Meditation.

So too in genAI image creations, it is the viewer who imposes an origin-story to the output, a reason for that particular collection of pixels. This imposition is a counter-authenticity, or an inauthenticity but, again, not inauthenticity as lack, but as potential.

## 6 Inauthenticity as a creativity engine in generative AI

The purpose of this paper is to redeem the inauthenticity of AI creativity. It is to convert inauthenticity into a reason to want artificial creativity, even a reason to want artificiality more than traditional performances and productions. There are two arguments.

## 6.1 Argument 1

Authentic art is confining; inauthentic AI creativity is liberating. The value of Benjamin's aura is the control that original artists can exert over their work, even across centuries. The inauthentic art of the digital shine makes no controlling claims. Viewers are free to create any imaginable origin-story for their genAI product, and align it with any possible future project.

In discussions of art and music today, we sometimes hear of cultural appropriation and sense the term as a threat or barrier: people who are not authentic representatives of one or another tradition or culture or music may not reproduce or play it. For instance, like many white musicians Justin Bieber has been accused of exploiting African-American musical patterns and, in Bieber's case, also of twisting his hair into dreadlocks without properly acknowledging the cultural origin of the style. Reasonable people disagree about the line between exploitation and appreciative influence, but what matters here is that the entire discussion evaporates amid genAI production. Cultural and political limits that may constrain creativity fall away entirely with large genAI platforms because there is no original culture to be offended. Dilemmas about debts earlier creators are erased because there is no founding artist to be disrespected.

A clear example of liberation in inauthenticity can be found in creative co-productions between humans and AI, and by contrasting them with co-production between humans.

Starting with humans working together, the limitations of co-productions are perceivable in musical jam sessions, in performances where the players must defer to each other as one and then another takes the lead role. Especially in jazz and instrumental improvisations, one player picks up the sound that another is putting down, and then steps forward to take the lead and add their own twist to the notes before leaving the result for the next musician to continue. For the performance to advance, each sequential artist must respect the style of the musician who came before, as well as the particular sound of the player who comes next. Everyone must conform to the co-creators around them. Of course there is an interpersonal magic here, and a kind of communal joy, but there is also a restriction on creativity: no one person can take the sound entirely in their own direction.

This limit is overcome when a musician plays with an AI partner. The deference, the respect that hinders individual creativity disappears because no one cares how a human leads the machine. There are no objections when the human player demands that the machine assume the subservient role of constantly adjusting its musical production to abjectly following the human lead. This liberation is audible in a widely cited experiment in improvisation between human players and an AI drummer [5]. In the experiment, titled "In a silent way: Communication between ai and improvising musicians beyond sound", an AI is trained to provide a rhythmic supplement for a leading horn player. The AI, in other words, is an improvisation partner. Ethically, the critical element is that the human leader is completely free to direct the music, and also to demand that the AI partner serves for inspiration but never as a limitation. The AI partner has its own agency in the sense that it plays music in (artificially) original ways, but never in a way that challenges or restrains human creativity by ruling that some musical directions are out of bounds, culturally insensitive, or impermissible for reason of conflicting with the authentic identity of the AI composer.

When this happens – when authenticity is erased - creativity maximizes because there is no need for one artist to restrain their own skill to accommodate another. Stronger, the deference is impossible because there is nothing to defer to, no origin-story, no cultural and historical source. The AI provides inspiration for unlimited human originality.

### 6.2 Argument 2

Authentic art imbued with an aura proposes to memorialize a way of life, it captures something of how the artist lived. The opportunity presented by inauthentic creativity is distinct, it is to create new ways of life. Artificial creativity stimulates creation on the human level by opening possibilities for artists to become different from who they have been. In other words, it is not just that artificial creativity is liberating for art fabrication, but also for life construction.

Part of AI art is the generic sheen, the sense that it does not belong anywhere, or to anyone: it portrays human experience without any specific human's experience. One derivative of this reality is the rendering of possibilities for human life as literally endless: creativity unbound from any specific experience is open to every experience.

As a rudimentary example, there is Max Hawkins, the software engineer who devised an app for Facebook that short-circuited the platform's recommendation algorithm and instead proposed random events and meetings. So, instead of the usual suggestions crafted to reinforce his existing habits and preferences, he got sent in tangential directions that split away from his current life. He ended up visiting picnics for Russian immigrants and similar, even though he was neither Russian nor an immigrant.

The point is that creative experiences detached from the authentic self can lead to divergent projects, ones that break away from established precedents and identities. Instead of reinforcing the authentic person who is a user on the Facebook platform, the recommendations inspire users to try new diets, enter unfamiliar social circles, entertain unpracticed habits. Inauthentic recommendations generate the kinds of experiences that can emerge when we no longer ask where someone came from, and when we no longer worry about the legitimacy of any claimed heritage or identity. Instead, we ask about disruptive possibilities, and pursue them when they arise.

This project in inauthenticity has been explored on the legal level as the right to discontinuity [6], and also on the ethical level as genhumanism [7]. In both cases, what is being disputed is the traditional artistic idea that there is value in being true to who you are. The fundamental claim of artificial creativity and art is that there is value in not being true to who you are (the claim is even radical as there is no one to be true to). And, if that is right, then inauthenticity in creativity – the absence of any origin story preceding AI outputs – is no longer deficiency, but potential.

The task that remains is to investigate exactly how the artistic creativity of genAI can be converted into humans developing new interests, tastes, preferences, and aspirations. Is it possible for artists to use their creations to change who they are instead of express who they are? Can the inauthenticity at the heart of AI creativity become an engine for identity transformation?

The answers are currently uncertain, but we do know that, for Benjamin, it was the withdrawal of authenticity under the pressure of mechanical reproduction that allowed us to see in isolation the aura of the artwork. The aura was always there, of course, but it was finally perceived as the sensation of its deflation and loss.

Today, the disappearance of authenticity under the pressure of generative AI has a different effect. Instead of revealing something as withdrawing from art, it exposes what *propels* a new kind of art. It is precisely the absence of any origin that maximizes a technological kind of creativity with ultimate effects for art and for humans that have not yet been contained or adequately described, even though they are already happening.

## 7 Conclusion

In this paper we proposed the concept of artificial creativity as a challenge to Benjamin's notion of an "aura" in art. While Benjamin considers authenticity to be a fundamental aspect of art, here we have argued that generative artificial intelligence embodies pure inauthenticity. Two characteristics of purely inauthentic art have been discussed: elusiveness and reflection. Elusiveness refers to the difficulty in identifying an origin or backstory for the artwork, whereas reflection refers to the ability for perceivers to impose any origin that serves their own purposes. The paper has argued that these traits expand the boundaries of artistic and creative possibilities. To exemplify this, the discussion has included a scenario involving musical improvisation with an artificial intelligence collaborator. Finally, the paper has questioned whether the inauthentic creativity seen in AI-generated art can be applied to human experiences and extended to our sense of our identities.